\documentclass[journal,lettersize]{IEEEtran}
\usepackage{amsmath,amsfonts}
\usepackage{algorithmic}
\usepackage{array}
\usepackage{subfig}
\usepackage{textcomp}
\usepackage{stfloats}
\usepackage{url}
\usepackage{verbatim}
\usepackage{graphicx}
\usepackage{hyperref}
\def\BibTeX{{\rm B\kern-.05em{\sc i\kern-.025em b}\kern-.08em
    T\kern-.1667em\lower.7ex\hbox{E}\kern-.125emX}}
\usepackage{balance}
\hypersetup{draft}
\begin{document}

\title{Compression Ratio Learning and Semantic Communications for Video Imaging}
\author{Bowen Zhang\IEEEauthorrefmark{1}, Zhijin Qin\IEEEauthorrefmark{2}, and Geoffrey Ye Li\IEEEauthorrefmark{1}\\
\small \IEEEauthorrefmark{1} Department of Electrical and Electronic Engineering, Imperial College London, London, UK\\
\small \IEEEauthorrefmark{2} Department of Electronic Engineering, Tsinghua University, Beijing, China\\
\small \{k.zhang21, geoffrey.li\}@imperial.ac.uk, qinzhijin@tsinghua.edu.cn
}


\maketitle

\begin{abstract}
Camera sensors have been widely used in intelligent robotic systems. Developing camera sensors with high sensing efficiency has always been important to reduce the power, memory, and other related resources. Inspired by recent success on programmable sensors and deep optic methods, we design a novel video compressed sensing system with spatially-variant compression ratios, which achieves higher imaging quality than the existing snapshot compressed imaging methods with the same sensing costs. In this article, we also investigate the data transmission methods for programmable sensors, where the performance of communication systems is evaluated by the reconstructed images or videos rather than the transmission of sensor data itself. Usually, different reconstruction algorithms are designed for applications in high dynamic range imaging, video compressive sensing, or motion debluring. This task-aware property inspires a semantic communication framework for programmable sensors. In this work, a policy-gradient based reinforcement learning method is introduced to achieve the explicit trade-off between the compression (or transmission) rate and the image distortion. Numerical results show the superiority of the proposed methods over existing baselines. 
\end{abstract}


\section{Introduction}
Cameras have become ubiquitous in robotic systems and enable robots to evaluate the environment, infer their status, and make intelligent decisions. However, modern cameras always suffer from the age-old problem of limited dynamic range caused by the full-well capacity, dark current noise, and read noise \cite{debevec2023recovering}, and the challenge of motion blur resulting from either object motion during the image exposure or camera motions due to the robot movements. 

To overcome some of these challenges, computational imaging techniques have been widely used for generating high dynamic range (HDR) images \cite{hasinoff2016burst,nayar2000high,nayar2003adaptive} or high-speed videos \cite{gu2010coded, sankaranarayanan2010compressive}. These methods use spatially-varying pixel exposure or pixel-wise coded exposure along with optimization techniques to get improved performance in imaging systems. Among these methods, snapshot compressive imaging (SCI) \cite{llull2013coded, wagadarikar2008single, wagadarikar2009video} captures a set of consecutive video frames with one single exposure and is popular for realizing sparse measurements and low requirements on memory, bandwidth, and power \cite{yuan2020plug}. 

Due to the hardware limitation of conventional cameras, existing SCI systems capture an image by exposing photo-sensitive
elements for a fixed exposure time, leading to a fixed temporal compression ratio for all pixels. The fixed compression ratio of the current SCI systems severely limits their abilities to record natural scenes in a measurement-efficient way. With a small compression ratio, more measurements are sampled, increasing the power, memory, and related-resources. With a large compression ratio, fewer samples are needed but the reconstructed video frames have poor quality. If pixels within SCI systems can be generated under different compression ratios, the video compressive sensing system can maintain high quality reconstruction and achieve efficient measurement. However, such a video compressive sensing system not only has special requirements on the hardware, but also requires an optimal pixel-wise compression ratio assignment policy, which is not trivial as both the shot/read noises and object/camera motions will affect the choice of compression ratios.  

Fortunately, programmable sensors or focal-plane sensor–processors \cite{carey2013100,chen2017feature,wong2020analognet} can vary compression ratios spatially through pixel-level control of exposure time and read-out operations. Recent works in deep optics, on the other hand, demonstrate the superiority of jointly learning optic parameters and image processing methods over the traditional computational imaging techniques with heuristic designs on optics for applications
in HDR imaging \cite{metzler2020deep}, video compressive sensing \cite{martel2020neural,iliadis2020deepbinarymask}\footnote{In these deep optic-based works on video compressive sensing, the coded aperture and the exposure time for generating a snapshot image is optimized but the compression ratio, the number of measurements over a time window, is fixed. Different from these works, we focus on studying the shutter speed and read-out frequencies to adjust per-pixel compression ratio. The most significant difference is that some pixel locations will have more measurements than other locations in the captured sensor data of our system.}, and motion deblurring \cite{nguyen2022learning}. Inspired by these pioneering works, we focus on minimizing the number of measurements for various applications and develop a novel video compressive sensing system with pixel-wise compression ratios, where the ratio allocation policy and the video reconstruction algorithms are optimized through a combination of supervised learning and policy gradient reinforcement learning (RL) \cite{sutton1999policy, liu2017improved}.  

In addition to sensor development, we also study the optimal transmission method for sensor data generated by programmable sensors in terms of source and channel coding methods since the raw sensor data collected by the sensors at the robots need to be sent to the admins for human-robot interaction. For the sensing systems with image-type data \cite{nguyen2022learning,martel2020neural,vargas2021time}, off-the-shelf codecs (e.g. BPG \cite{bpg2017} or JPEG \cite{wallace1991jpeg}) as well as
other deep image compression methods \cite{balle2017end} can be used as potential source coding and the fifth‐generation (5G)-standardized LDPC \cite{thorpe2003low} or polar codes \cite{vangala2015comparative} can be chosen for channel coding. However, it is sub-optimal to design the source and channel coding for sensor data independently from the deep optics and video reconstruction algorithms for the following two-fold reasons. The distribution of sensor data will be different from that of regular images under the influence of deep optics. Different pixels in the sensor data contribute differently to the following video reconstruction process and should be compressed differently. 

To address these challenges, semantic communications \cite{qin2021semantic, 10198383, lan2021semantic,jankowski2020wireless} are promising solutions. The deep optics and video reconstruction networks define special message generation and interpretation processes between transceivers, which are called semantic encoders and decoders in semantic communications, respectively. The communication systems should be optimized to ensure the interpretations of the messages through semantic decoders are correct rather than the delivery of raw sensor data itself. To realize semantic communications, one potential way is to introduce task-aware compression \cite{song2021variable} for programmable sensors and design the compression methods under the guidance of the video reconstruction process. In this case, the channel coding is still designed separately while earlier studies on joint source and channel coding (JSCC) \cite{bourtsoulatze2019deep} have demonstrated the benefits of co-designing the source and channel coding process. In this work, a semantic communication framework that achieves the joint optimization of deep optics, data compression, channel coding, and reconstruction is designed, which significantly improves the transmission efficiency of sensor data.  

Specifically, our contributions can be summarized as follows,
\begin{itemize}
\item We introduce a RL-based method for adjusting the parameters in programmable sensors, which differs from existing deep optic methods based on differentiable models or functions \cite{nguyen2022learning}. 
\item We build a novel video compressive sensing system with spatially-variant compression ratios, where the ratio allocation policy is learned through an explicit rate-distortion function. 
\item We introduce a RL-based method for the explicit trade-off between transmission rates and task accuracy in semantic communications, where the rate allocation policy is trained jointly with coding modules. 
\item We propose a semantic communication system for programmable sensors, realizing the co-design of deep optics, data compression, channel coding, and reconstruction algorithms. 
\end{itemize}

\section{Video Compressed Sensing with Spatially-variant Ratios}
\label{sec:sense}

\begin{figure}[ht]
\centering
\includegraphics[scale=0.32]{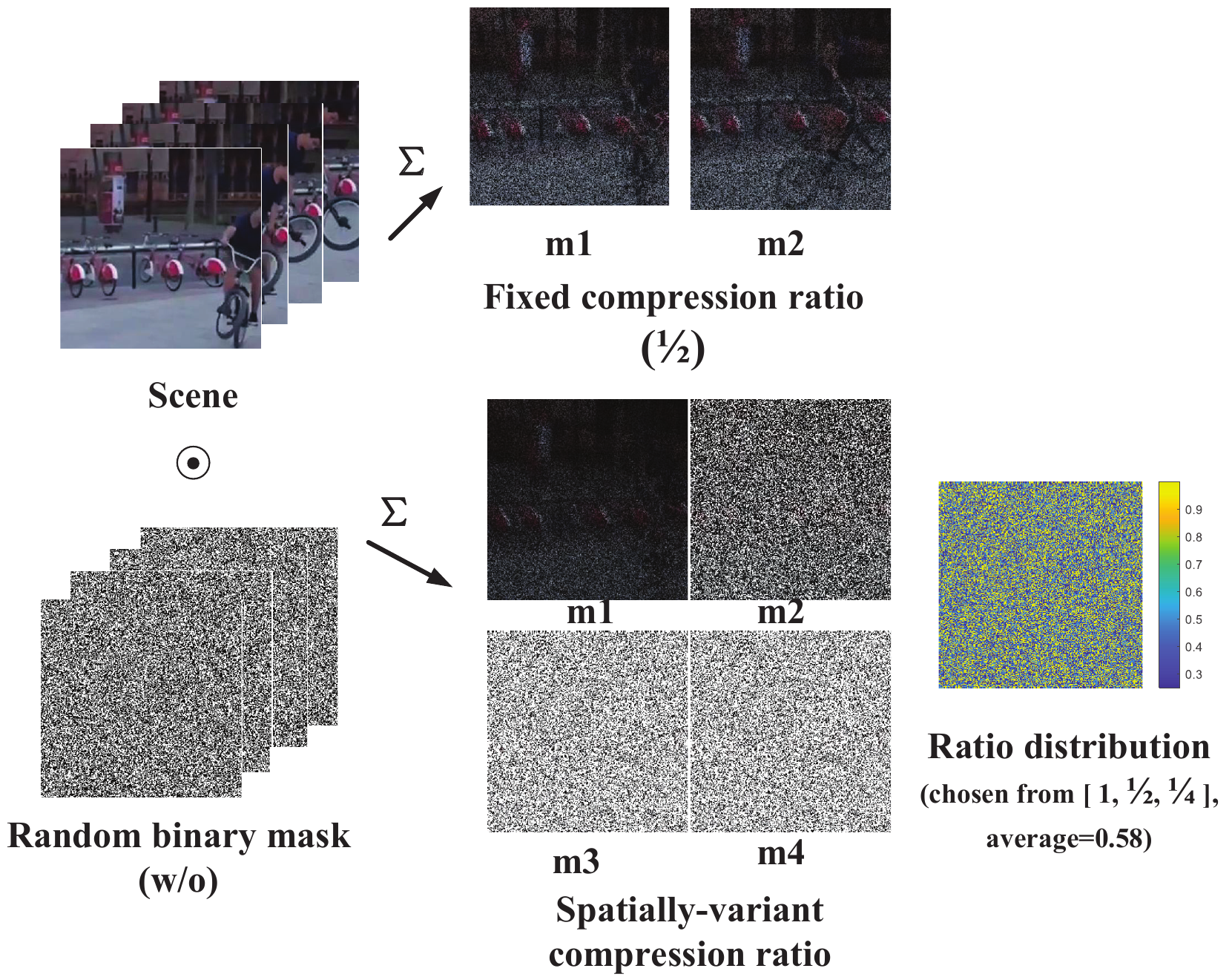}
\caption{The illustration of spatially-variant compression ratios in video compressed sensing.}
\label{fig:basic_idea}
\end{figure}
As shown in Fig. \ref{fig:basic_idea}, the principle of the proposed video compressed sensing system is to learn spatially-variant compression ratios for better imaging quality but with a fewer number of measurements. This idea can be further combined with the existing deep optic-based SCI systems with the pixel-wise coded aperture and shuttering functions \cite{vargas2021time}. In SCI systems with a fixed compression ratio, a fixed number of measurements will be generated for all spatial locations. While in the proposed system with spatially-variant ratios, some locations will have more measurements than others. For example, given four video frames, two snapshots (m1, m2) will be captured in the current SCI systems with $1/2$ ratio. By contrast, the proposed systems will generate one snapshot with dense measurements (m1) and three other snapshots with sparse measurements (m2, m3, m4). Specifically, spatial locations with $1/4$ ratio will only have measurements on m1 after one readout operation while locations with $1$ ratio will have values on all four snapshots after four readout operations. To improve the sensing efficiency, the ratio map should be jointly designed with the deep optics and video reconstruction algorithms. In this section, we will introduce the details of the proposed system including the forward model and the training losses. 

\begin{figure*}[t]
\centering
\includegraphics[scale=0.55]{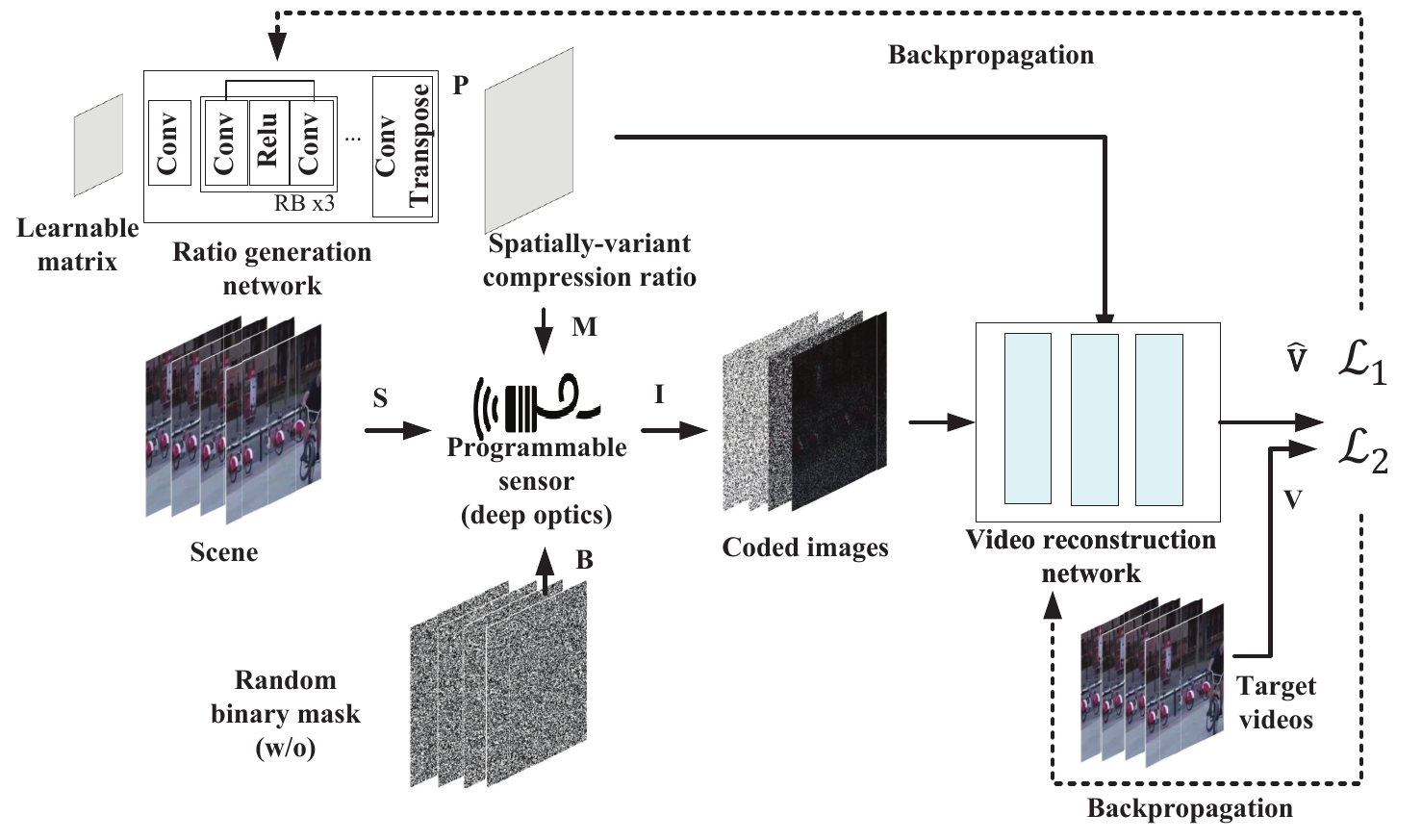}
\caption{The overview architecture of the proposed video compressed sensing system with spatially-variant ratios.}
\label{fig:sensing_framework}
\end{figure*}
\subsection{System overview}
We show the overall pipeline of the proposed sensing system in Fig \ref{fig:sensing_framework}. Denote $H$ and $W$ as the height and width of video frames. We first generate a compression ratio map, $M\in \mathcal{R}^{H\times W \times 1}$, from a small trainable matrix using a ratio generation network with one convolution layer, three residual blocks, and one transposed convolution layer. The spatial size of the trainable matrix is set as 1/8 of the ratio map. We then simulate the capture of a scene, $S\in \mathcal{R}^{H\times W \times T}$, with a programmable sensor using $M$ and possibly an extra random binary mask, $B\in \mathcal{R}^{H\times W \times T}$, which is referred to as coded apertures in earlier works \cite{vargas2021time, yuan2020plug}. The captured sensor data, $I$, stacked with $M$ is then fed into a video reconstruction network to produce a reconstructed video, $\hat{V}\in \mathcal{R}^{H\times W \times T}$. Two training losses ($\mathcal{L}_{1}$, $\mathcal{L}_{2}$) are designed to update the learnable matrix, the ratio generation network, and the video reconstruction network to improve the reconstruction quality while at the same time reduce the average compression ratio.  

\begin{figure}[t]
\centering
\includegraphics[scale=0.95]{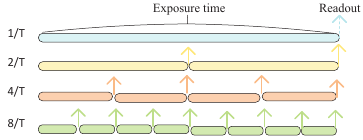}
\caption{The exposure time and readout operations in different compression ratios.}
\label{fig:shutter_function}
\end{figure}

\subsection{Compression ratio generation}
\label{sec:ratio_generation}
Given $T$ video frames, five kinds of compression ratios are considered during the ratio map design, i.e. $0$, $1/T$, $2/T$, $4/T$, $8/T$. Therefore, each pixel in $M$ takes one of the five discrete values. To generate the ratio map, the feature maps generated by the ratio generation network are designed to have five channels. After applying a Softmax function to the channel dimension of the feature maps, each channel indicates the discrete possibility of taking the corresponding ratio. Denote the feature maps representing the discrete probability distributions as $P \in \mathcal{R}^{H\times W \times 5}$. The final ratio map is obtained through a sampling operation according to $P$.

\subsection{Sensor forward model with varying ratios}
\label{sec:forward}
The learned ratio map will guide the behaviors of the programmable sensor. The exposure time and readout operations under different ratios are summarized in Fig. \ref{fig:shutter_function}. We consider that each measurement is read out after a continuous exposure ends and before a new round of exposure begins. Also, the generated "signal" electrons are cleared after each read-out operation. For example, when the ratio is $2/T$, each exposure lasts $T/2$ frame time, generating $2$ measurements in total. In each exposure, the signal will integrate from 0. For a specific ratio, $r$, the equivalent measurement matrix, $A_{r}$, is of size $rT \times T$ with each row representing one measurement base. For the $i$-th ($i=1,2,..,rT$) row, $[A_{r}]_{i:}$ is a binary vector with the $(i-1)/r$ column to the $i/r$ column taking values 1 while for other columns 0. For example. $A_{2/T}=\begin{bmatrix} 
1&\cdots&1&1&0&\cdots&0&0\\
0&\cdots&0&0&1&\cdots&1&1
\end{bmatrix}$. The first row indicates the first measurement is the integrated signal from frame time $1$ to $T/2$ while for the second row indicating the second measurement from time $T/2+1$ to $T$. With the measurement matrices, the sensing process at spatial location $p$ with $r_{p}$ ratio can be modelled as,
\begin{equation}
I_{p}=\mathcal{U}(A_{r_{p}}S_{p}),
\label{equ:1}
\end{equation}
where $S_{p}\in \mathcal{R}^{T\times 1}$ is the signal at spatial location $p$, $I_{p} \in \mathcal{R}^{r_{p}T\times 1}$ is the captured sensor data at location $p$, $\mathcal{U}(\cdot)$ represents the camera exposure function capturing
the noise effects of the camera \cite{nguyen2022learning}.  

If the random binary masks, $B \in \{0,1\}^{H\times W \times T}$ are used to modulate the signal \cite{vargas2021time}, the sensing process can be modelled as,
\begin{equation}
I_{p}=\mathcal{U}(A_{r_{p}}(S_{p}\cdot B_{p})),
\label{equ:2}
\end{equation}
where $\cdot$ represents element-wise multiplication, $B_{p}\in \mathcal{R}^{T\times 1}$ is the signal at spatial location $p$.

Following \cite{nguyen2022learning}, we consider two kinds of noises, i.e. shot noise $n_{s}\in \mathcal{N}(0,\sigma_{s})$ and read noise $n_{r}\in \mathcal{N}(0,\sigma_{r})$ in the camera exposure function. The level of shot noise is proportional to the signal electrons' strength. For a signal $e$ in [0,1] range, $\sigma_{s}=\sqrt{e}\sigma_{ss}$, where $\sigma_{ss}$ is independent of signals. By contrast, the level of read noise is fixed for a camera, depending on the photon flux. We use the same settings for $\sigma_{ss}$ and $\sigma_{r}$ as \cite{nguyen2022learning}, that is, $\sigma_{ss}=4.95\times 10^{-3}$ and $\sigma_{r}=7.25\times 10^{-3}$. With the definition of noises, $\mathcal{U}(e)=e+n_{s}+n_{r}$.

Note that increasing $r$ will decrease the signal strength of each measurement because of the short exposure. Since the distortion effects of shot noise and read noise on signals will increase as the signal strength decreases, the imaging quality will not increase unlimitedly as $r$ increases. Also, using a long exposure will be more beneficial for stationary scenes as the signal strength can increase but not good for moving objects. The best choice of ratios should be a mix of small and large compression ratios for different locations so that the captured sensor data can take benefits from both the long exposure with a larger signal-to-noise ratio and the short exposure with less motion blur.  

\begin{figure*}[ht]
\centering
\includegraphics[scale=0.4]{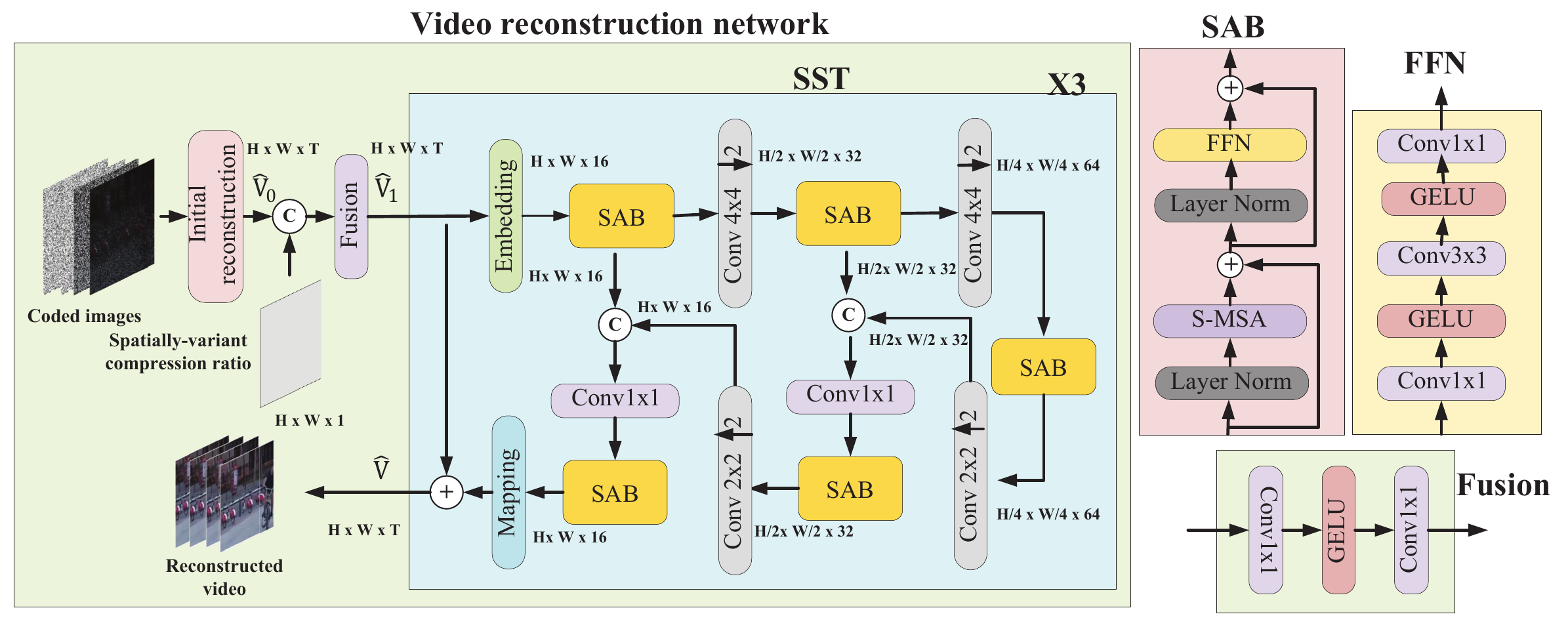}
\caption{The architectural details of video reconstruction network.}
\label{fig:video_re}
\end{figure*}

\subsection{Video reconstruction network}
After obtaining $I$, the next step is to reconstruct the targeted video using a deep neural network. The overall architecture of the proposed video reconstruction network is shown in Fig. \ref{fig:video_re}. Our network consists of three components,  an initial reconstruction stage (IR), a fusion network (FN), and a deep reconstruction network (DRN) built based on the  single-stage spectral-wise transformers (SST) proposed in \cite{cai2022multi}. The IR and FN are introduced to mitigate the influence of spatially-variant $M$ and $B$. Specifically, if $B$ is not considered, we get the initial reconstruction, $\hat{V}_{0} \in \mathcal{R}^{H\times W \times T}$, in IR by considering Eq. (\ref{equ:1}),
\begin{equation}
\hat{V}_{0_{p}}=A_{r_{p}}^{T}(A_{r_{p}}A_{r_{p}}^{T})^{-1}(I_{p}),
\label{equ:3}
\end{equation}
where $(A_{r_{p}}A_{r_{p}}^{T})^{-1}$ is fortunately a diagonal matrix as defined in Sec. \ref{sec:forward}. By contrast, if $B$ is considered, we first rewrite Eq. (\ref{equ:2}) as,
\begin{equation}
I_{p}=\mathcal{U}(\hat{A}_{r_{p}}S_{p}),
\label{equ:4}
\end{equation}
where $\hat{A}_{r_{p}}=A_{r_{p}}\text{diag}(B_{p})$ and $\text{diag}(B_{p})$ denotes the $T\times T$ diagonal matrix constructed from $B_{p}$. In this case, some rows of $\hat{A}_{r_{p}}$ may be all zero, making $\hat{A}_{r_{p}}\hat{A}_{r_{p}}^{T}$ non-invertible. Since the row vectors of $\hat{A}_{r_{p}}$ are orthogonal to each other, we reconstruct $\hat{V}_{0}$ by using the non-zero row vectors of $\hat{A}_{r_{p}}$ independently,
\begin{equation}
\hat{V}_{0_{p}}=\sum_{\substack{i=1 \\ [\hat{A}_{r_{p}}]_{i:}\neq \mathbf{0}_{1\times T}}}^{rT}[\hat{A}_{r_{p}}]_{i:}^{T}([\hat{A}_{r_{p}}]_{i:}[\hat{A}_{r_{p}}]_{i:}^{T})^{-1}[I_{p}]_{i:}+ \mathbf{0}_{T\times 1}.
\label{equ:5}
\end{equation}

Following the coarse-to-fine criteria, we further use a shallow FN to deal with the spatially-variant coded exposure and ratios, which takes $\hat{V}_{0_{p}}$, $M$ and possibly $B$ as inputs and outputs the second-level reconstruction results, $\hat{V}_{1} \in \mathcal{R}^{H\times W \times T}$. The architecture of FN is shown in Fig \ref{fig:video_re}.

Next, as illustrated in Fig \ref{fig:video_re}, the DRN takes $\hat{V}_{1}$ as the input and reconstructs the final videos, $\hat{V}$, by cascading three SSTs \cite{cai2022multi}. Each SST follows the design of U-net \cite{ronneberger2015u} with an encoder, a bottleneck, and a decoder. The embedding and mapping block are convolutional layers (conv) with 3×3 kernels. The feature maps in the encoder sequentially pass through one spectral-wise attention block (SAB), one conv with stride 2 (for downsampling), one SAB and one conv with stride 2. The bottleneck is one SAB. The decoder has a symmetrical architecture to the encoder. Following the spirit of U-Net, the skip connections are used for feature aggregation between the encoder and decoder to alleviate the information loss from the downsampling operation. The basic unit of SST is SAB, whose architecture is also shown in Fig \ref{fig:video_re}. It has one feed forward network (FFN), one spectral-wise multi-head self-attention (S-MSA), and two layer-normalization. Different from original MSA that calculates the self-attention along the spatial dimension, S-MSA regards each feature map as a token and calculates the self-attention along the
channel dimension, making it computational-effective. More details of S-MSA are explained in \cite{cai2022multi} \footnote{As the network design of DRN is not the main focus of the paper, we omit some details here for page limits. Interesting readers can refer to the original paper for more details.}.  

\subsection{Training losses}
\label{sec:loss_sensing}
We use $\mathcal{L}_{1}$, $\mathcal{L}_{2}$ to train the ratio generation parts and the video reconstruction network, respectively. Specifically, the learnable matrix and the ratio generation network are trained based on policy gradients. In our system, each spatial location in $P$ is regarded as an agent and its action space is the available compression ratios. Earlier works have proved the global convergence of policy gradient RL in multi-agency situation \cite{leonardos2021global}. We define the reward of each location, $Q_{p}(r_{p})$, under action $r_{p}$ according to the rate-distortion trade-off theory,
\begin{equation}
Q_{p}(r_{p})=\log(1/||V_{p}-\hat{V}_{p}||^{2})-\lambda r_{p}T,
\label{equ:6}
\end{equation}
where $V$ denotes target videos, $||V_{p}-\hat{V}_{p}||^{2}$ denotes the mean-squared error (MSE) (or called distortion) of the reconstructed videos at spatial location $p$, $r_{p}T$ denotes the number of measurements employed at location $p$, which can also be viewed as the compression rate, $\lambda$ is an introduced parameters for rate-distortion trade-off. Increasing $\lambda$ will penalize more on the compression rate, leading to a smaller average compression ratio. As this is not a sequential decision problem, there is no need to define future rewards. Although $||V_{p}-\hat{V}_{p}||^{2}$ can only approximately evaluate the effect of action $r_{p}$ on the $V_{p}$ as the DRN will aggregate information from neighbouring pixels, it is also the most direct way to evaluate action $r_{p}$. 

At the same time, the expected reward, $J_{p}$, is the value function for spatial location $p$, where the expectation is w.r.t. $r_{p}$ with probability $P_{p}$. Note that, $\text{Min} \mathcal{L}_{1}=\text{Max}\sum_{p} J_{p}$. Following \cite{liu2017improved},  we can approximate the gradient of the $J_{p}$ to parameters $\theta$ in ratio generation parts with samples generated from $P_{p}$,
\begin{equation}
\nabla_{\theta}J_{p}=\mathbb{E}_{r_{p}}\nabla_{\theta}\log P_{p}(r_{p}) \times Q_{p}(r_{p}),
\label{equ:7}
\end{equation}
where $P_{p}(r_{p})$ denotes the probability of chosen action $r_{p}$ from distribution $P_{p}$. Note that the update of $\theta$ is based on the average value of $\nabla_{\theta}J_{p}$ from different $p$. 

On the other hand, the video reconstruction network is trained in a supervised way based on the MSE between $V$ and $\hat{V}$,
\begin{equation}
\mathcal{L}_{2}=||V-\hat{V}||^{2},
\label{equ:8}
\end{equation}

\section{Efficient transmission of data from programmable sensors}
\label{sec:transmission}
Sometimes the sensor data captured by robots' on-device sensors need to be sent to remote servers/devices through wireless channels for data storage or video reconstruction. Efficient transmission of these sensor data requires advanced data compression techniques to reduce the data rates and bandwidth requirement. As discussed above, existing communication systems focusing on reproducing the raw sensor data are sub-optimal. As a part of semantic communications, designing the compression methods concerning the video reconstruction network based on task-aware compression \cite{song2021variable} is promising. After compressing the original sensor data into compact bit streams, the transmitter will add parity bits and modulate the streams for robust transmission over unreliable channels using off-the-shelf methods. The number of modulated symbols after channel coding and modulation is the communication cost. 

Nevertheless, there is an increasing belief in the communication community that the classic framework based on the Shannon separation theory needs to be upgraded for joint designs \cite{jankowski2020wireless}. For semantic communication systems, jointly optimizing the channel coding and modulation with the other components may lead to better video reconstruction quality with fewer communication resources. 
 
In this section, we will design semantic communication frameworks for the proposed video compressed sensing systems based on both the task-aware compression and semantic communications with joint designs. 

\begin{figure*}[t]
\centering
\includegraphics[scale=0.41]{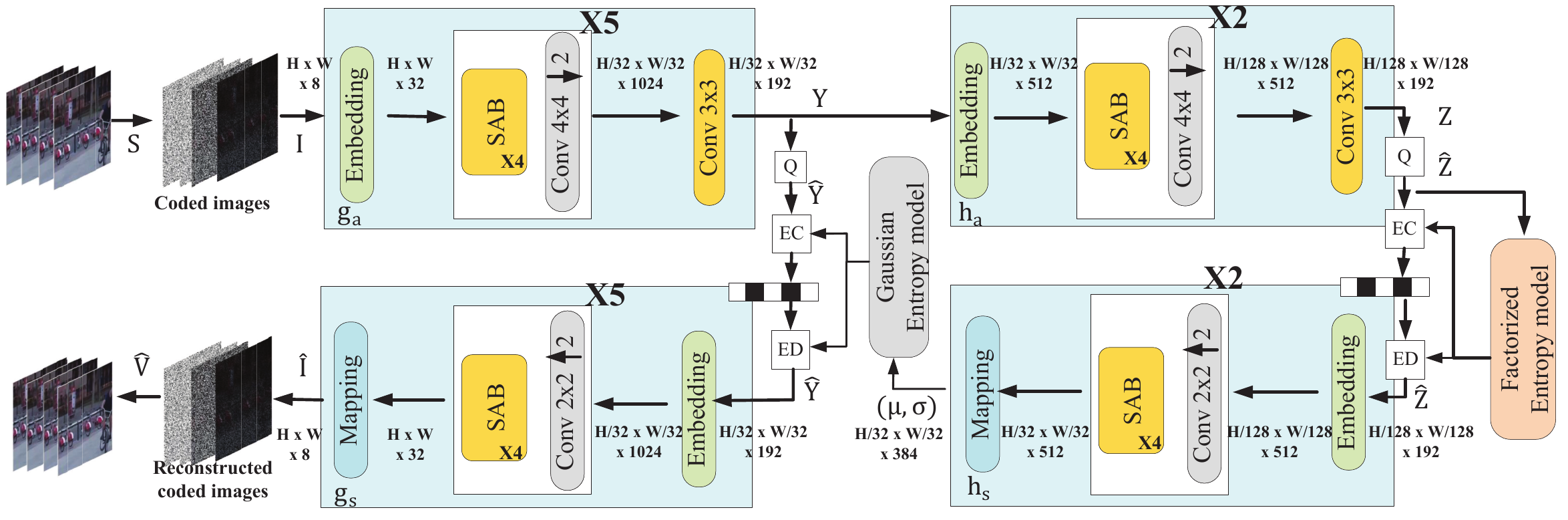}
\caption{The overall framework of task-aware compression methods.}
\label{fig:task-compress}
\end{figure*}

\subsection{Task-aware compression}
\subsubsection{Architecture} The overall architecture of the deployed deep compression methods for the proposed video compressed sensing systems is shown in Fig. \ref{fig:task-compress}. It is mainly based on the architecture used in \cite{song2021variable} by substituting the basic feature extraction unit from convolutional blocks to SABs. Some other changes are also made for simplicity\footnote{As the learned compression ratio is universal to all scenes, the spatial importance of $I$ should also be independent from the contents of $I$, therefore, the side links conditioned on a quality map used in \cite{song2021variable} can be safely removed. Also, instead of estimating the quality map by a task loss function, we directly change the the distortion
metric in the training loss of \cite{song2021variable} to task loss, which should be more precise in describing the spatial importance.}. For a sensor data $I$, we first reshape its dimension to $H\times W\times 8$, as each spatial location has at most eight measurements. Zero-padding is used for locations with fewer measurements. An encoder, $g_{a}$, then takes the reshaped sensor data as inputs and generates a latent representation $Y \in \mathcal{R}^{\frac{H}{32}\times \frac{W}{32} \times 192}$, which is given by, $Y=g_{a}(I)$. $Y$ is then quantized to $\hat{Y}$ by a quantizer. The embedding and mapping blocks are conv3x3. The features inside the encoder subsequently go through five SAB$\times 4$-downsampling$\times 1$ pairs. Each downsampling operation will decrease the spatial size of the features by $\frac{1}{4}$ but double the channel dimension. Next, a hyper-encoder, $h_{a}$, takes $Y$ as inputs and generates an image-specific side information $Z \in \mathcal{R}^{\frac{H}{128}\times \frac{W}{128} \times 192}$ via $Z=h_{a}(Y)$. The channel dimension of the features keeps constant in $h_{a}$. Then the quantized side information
$\hat{Z}= Q(z)$ is saved as a lossless bitstream through a factorized entropy model and entropy coding. After that $\hat{Z}$ is forwarded to the hyper-decoder, $h_{s}$, to draw the parameters $(\mu, \sigma)$ of a Gaussian entropy model, which
approximates the distribution of $\hat{Y}$ and is used to save $\hat{Y}$ as a lossless bitstream.  For reconstructing $I$, a decoder, $g_{s}$, operates on $\hat{Y}$ and generates $I$ by $I=g_{s}(\hat{Y})$. At last, the reconstructed $\hat{I}$ is used for video reconstruction. Note that $g_{s}$ and $g_{a}$, $h_{s}$ and $h_{a}$ have symmetric architectures, respectively. 

\subsubsection{Training losses} The goal of task-aware compression is to minimize the length of the bitstreams and the distortion between $V$ and $\hat{V}$. This objective raises an optimization problem of minimizing $||V-\hat{V}||^{2}+\beta(-\text{log}_{2}P_{\hat{Y}}-\text{log}_{2}P_{\hat{Z}})$, where $||V-\hat{V}||^{2}$ denotes the video reconstruction quality, $-\text{log}_{2}P_{\hat{Y}}-\text{log}_{2}P_{\hat{Z}}$ represents the number of bits used to encode $\hat{Y}$ and $\hat{Z}$, and $\beta$ is an introduced trade-off parameter between coding rate and video distortion. Increasing $\beta$ will increase the compression ratio but decrease the video reconstruction quality. 

\subsubsection{Communication costs} After obtaining the compression systems with different compression ratios, we can then estimate the required number of modulated symbols to transmit these bit streams under different channel conditions. Specifically, the number of modulated symbols (in complex number) used for transmitting a bitstream depends on the implemented channel coding rate $r_c$ (\textit{e}.\textit{g}. 1/3, 1/2, 2/3) and modulation order $r_{m}$ (\textit{e}.\textit{g}. 4, 16, 64). Suppose the length of bitstream is $l_{b}$, the length of modulated symbols can be calculated as $l_{s}=\frac{l_{b}}{r_{c}\log_{2}(r_{M})}$. Another way to evaluate the communication costs is to use Shannon capacity theorem,  $l_{s}=\frac{l_{b}}{\log_{2}(1+snr)}$ if the signal-to-noise (SNR) ratio is $snr$.

\begin{figure*}[t]
\centering
\includegraphics[scale=0.4]{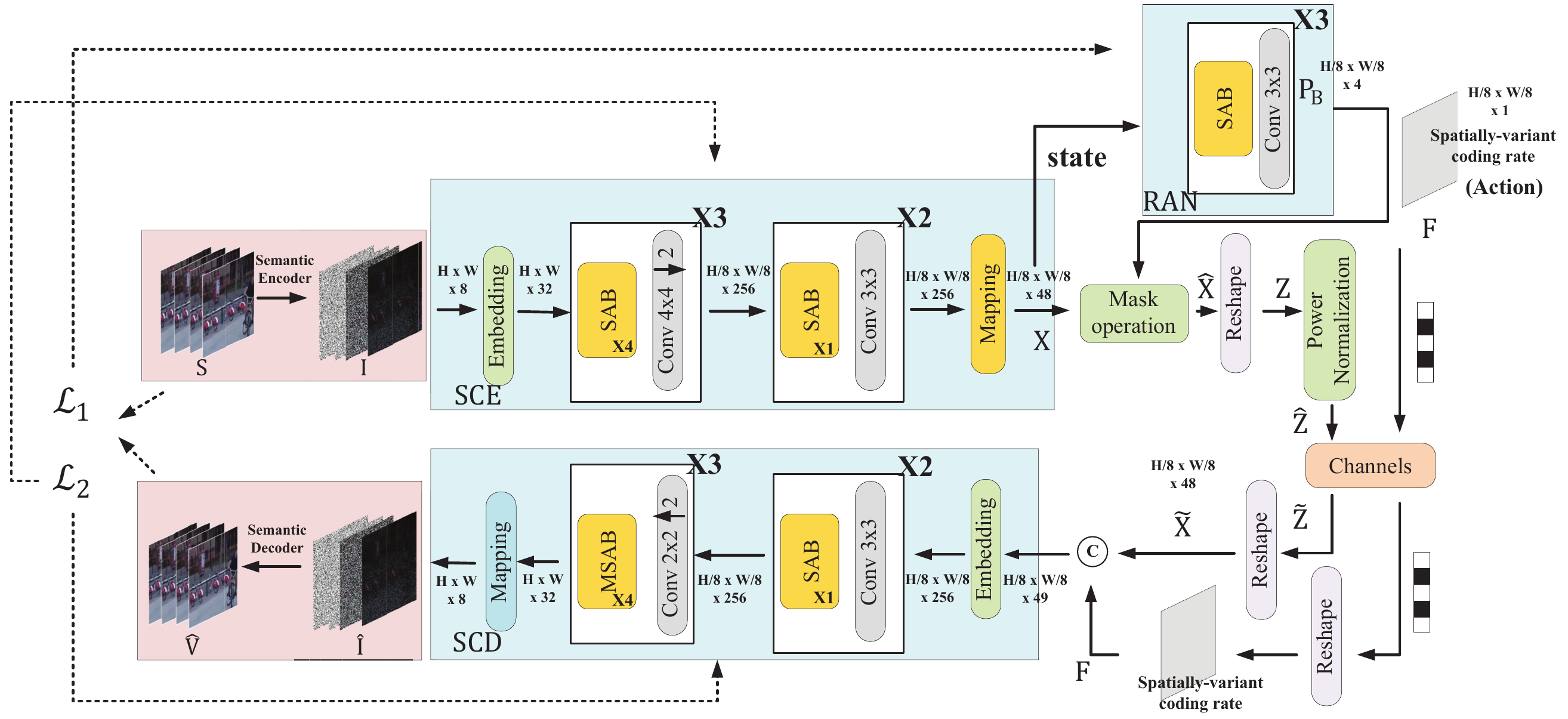}
\caption{The overall framework of semantic communications with joint designs. }
\label{fig:sem}
\end{figure*}

\subsection{Semantic communications with joint design} 
Although the aforementioned deep compression method enables the co-design of deep optics, algorithms, and source coding, the widely-used entropy coding method \cite{song2021variable, balle2017end} prohibits the co-design of communication-specific components with deep compression methods. When entropy coding methods are used, each bit from entropy coding methods needs to be transmitted error-freely; otherwise, error propagation will happen in the entropy decoding process. This property leaves the communication systems with no choice but treat each bit equally and carefully. 
Semantic communications with joint design, on the other hand, allow the effect of channel noise to be considered during the compression process. 

\subsubsection{Architecture} As shown in Fig. \ref{fig:sem}, the framework consists of three components: semantic coders, semantic-channel coders, and a rate allocation network (RAN). Semantic coders define a special message generation and interpretation method between transceivers based on a shared knowledge base. Semantic-channel coders directly learn the end-to-end mappings between semantic messages and modulated symbols. RAN is responsible for controlling the transmission rates. Different from deep compression methods that focus on the source coding rate only, the transmission rates in semantic communications depend on source coding rate, channel coding rate, and modulation order.    

Specifically, the deep optic methods are special semantic encoders, which encode natural scene $S$ into sensor data $I$ in a predefined way and the video reconstruction networks, which decode the target video $V$ from $I$, are special semantic decoders. The sensor data, $I$, is the semantic message of a scene to be shared between transmitters and receivers. Note that conventional communication systems emphasized the accurate transmission of $I$. With the semantic decoder (defining $I \rightarrow V$), the semantic communication systems should be optimized to maximize the quality of $V$ under limited communication costs. Also, this is the first attempt to define the semantic coders in the generation process of source data.   

Given $I$, the transmitter will use a semantic-channel encoder (SCE) to generate a predefined maximum number of modulated symbols, $X\in \mathcal{R}^{\frac{H}{8}\times \frac{W}{8} \times 48}$, from which some symbols will be chosen to transmit $I$ through noisy channels by rate control techniques \footnote{Otherwise, if all the messages are transmitted with the same number of modulated symbols, it disobeys the Shannon source coding theory saying that the minimal possible expected length of codewords should be a function of the entropy of the input word. }. This process can be modelled as,
\begin{equation}
X=\text{SCE}(I),
\label{equ:9}
\end{equation}
where the number of modulated symbols is measured in real number. The SCE is composed of an embedding layer, three consecutive SAB$\times 4$-downsampling$\times 1$ pairs, two stacked SAB$\times 1$-3$\times 3$Conv$\times 1$ pairs, and a mapping layer. The size of feature maps after each operation is shown in Fig. \ref{fig:sem}.

To adjust the communication costs according to the semantic contents of message $I$, the RAN takes $X$ as inputs, which are not only generated modulated symbols but also the high-level representations of $I$ in feature space, and generates a spatially-variant coding rate map $F \in \mathcal{R}^{\frac{H}{8}\times \frac{W}{8} \times 1}$. Each element in the spatial location of $F$ indicates how many symbols out of the $48$ symbols in $X$ at the same spatial location will be kept for transmission. Specifically, $F$ takes four discreet value $f \in \{1,2,3,4\}$, and only the first $12f$ symbols of $X$ in the channel dimension will be transmitted. The generation of $F$ is similar to the generation of $M$ in Sec. \ref{sec:ratio_generation}: the RAN generates a feature map $P_{F} \in \mathcal{R}^{\frac{H}{8}\times \frac{W}{8} \times 4}$ representing the discreet probability distribution of taking each value from $\{1,2,3,4\}$ and a sampling operation is conducted to generate $F$. With $F$, a mask operation is conducted on $X$ to delete unnecessary symbols. The remaining symbols $\hat{X}$ are then reshaped into a vector $Z \in \mathcal{R}^{n_{z} \times 1}$, where $n_{z}$ denotes the number of remaining symbols. Next, power normalization is applied to $Z$ to satisfy power constraints,
\begin{equation}
\hat{Z}=Z/||Z||^{2}.
\label{equ:10}
\end{equation}
Finally, $\hat{Z}$ is directly transmitted through wireless environments. The effect of channel noise on $\hat{Z}$ can be represented as,
\begin{equation}
\Tilde{Z}=h\hat{Z}+n,
\label{equ:11}
\end{equation}
where $h \in \mathcal{C}\mathcal{N}(0,1)$ is multiplicative noise and $n\in \mathcal{N}(0,\sigma_{n}I_{n_{z} \times n_{z}})$ is additive noise. In additive white Gaussian noise (AWGN) channel, $h=1$ and the value of $\sigma_{n}$ depends on the current signal-to-noise ratio (SNR).
Simultaneously, each element of $F$ can be quantized using 2 bits and the bitstream from $F$ is also transmitted. Different from the transmission of $\hat{Z}$, the bitstream of $F$ is transmitted in an error-free way as in the deep compression methods because a minor error in $F$ will make it impossible to reshape the flatten vector $Z$ back to $\hat{X}$.  

At the receiver side, $\Tilde{X}$ and $F$ are reshaped from $\Tilde{Z}$ and bitstreams respectively. After concatenation, they are fed into a semantic-channel decoder (SCD), which has a symmetric architecture to the SCE. This process can be modelled as,
\begin{equation}
\Tilde{I}=\text{SCD}(\Tilde{X},F),
\end{equation}
The recovered sensor data $\Tilde{I}$ is used to generate videos using the video generation network.  

\subsubsection{Training losses}
Similar to Sec. \ref{sec:loss_sensing}, we use $\mathcal{L}_{1}$ and $\mathcal{L}_{2}$ to train the RAN and SCE/SCD, respectively. Specifically, the RAN is trained based on policy gradients RL, where $X$ is the state, $F$ describes the actions, and each spatial location of $F$ is an agent. Note that the spatial size of $\hat{V}$ is 8 times that of $F$ and $X$ so the action taken at each spatial location of $F$ will have a strong effect on the reconstruction quality of a $8\times 8$ area of $\hat{V}$. Considering this, we first define $U=(V-\hat{V})^{2}$ and apply a $8\times 8$ average pooling to $U$, obtaining $\Tilde{U}\in \mathcal{R}^{\frac{H}{8}\times \frac{W}{8} \times 1}$. We then define the reward of location $p$, $Q_{F}^{p}(f_{p})$ , under action $f_{p}$ as follows,
\begin{equation}
\begin{aligned}
Q_{F}^{p}(f_{p})=\text{log}(1/\Tilde{U}_{p})-\mu f_{p},
\label{equ:12}
\end{aligned}
\end{equation}
where $\text{log}(1/\Tilde{U}_{p})$ describes the distortion caused by the agent at spatial location $p$, $f_{p}$ is the transmission rate (also the action) at $p$, $\mu$ is an introduced parameter for rate-distortion trade-off. Increasing $\mu$ will penalize more on the transmission rate, leading to fewer modulated symbols to be transmitted. We adjust the communication costs by tuning $\mu$.

At the same time, the expected reward, $J_{F}^{p}$, is the value function for spatial location $p$. $\text{Min} \mathcal{L}_{1}=\text{Max} \sum_{p}J_{F}^{p}$ and the gradient of $J_{F}^{p}$ to the parameters $\delta$ in RAN can be approximated as follows,
\begin{equation}
\begin{aligned}
\nabla_{\delta}J_{F}^{p}=\mathbb{E}_{f}\nabla_{\delta}\log P_{F}^{p}(f) \times Q_{F}^{p}(f),
\label{equ:13}
\end{aligned}
\end{equation}
Note that the update of $\delta$ is based on the average value of $\nabla_{\delta}J_{Q}^{p}$ from different locations.

On the other hand, the SCE and SCD are trained in a supervised way based on the MSE between $V$ and $\hat{V}$,
\begin{equation}
\begin{aligned}
\mathcal{L}_{2}=||V-\hat{V}||^{2},
\label{equ:14}
\end{aligned}
\end{equation}

\subsubsection{Communication costs}
The communication costs consist of two parts: the transmission of $\hat{Z}$ and $B$. As $\hat{Z}$ is of shape $n_{z}\times 1$, the number of complex modulated symbols used to transmit $\hat{Z}$ can be denoted as $l_{s}^{(1)}=n_{z}/2$. On the other hand, $B$ is of shape $n_{B}=\frac{HW}{64}$ and each element can be represented by 2 bits, therefore the bitstream length from $B$ can be modelled as $l_{b}=2n_{B}$. If the channel coding rate is $r_c$ and the modulation order is $r_{m}$, the length of the modulated symbols for $B$ can be calculated as $l_{s}^{(2)}=\frac{l_{b}}{r_{c}\log_{2}(r_{M})}$. The total communication costs are $l_{s}^{(1)}+l_{s}^{(2)}$.

\section{Experiments}
In this section, we first demonstrate the superiority of the proposed video compressed sensing system in terms of sampling rate. Based on the developed sensors, we then evaluate the performance of different communication systems regarding communication costs.

\subsection{Video imaging with spatially-variant compression ratios}
 The following experiments are conducted to evaluate the effectiveness of the method proposed in Sec. \ref{sec:sense}.
\subsubsection{Dataset}
Following \cite{nguyen2022learning}, we use the Need for Speed (NfS) dataset \cite{kiani2017need} to train the network and evaluate its performance. The NfS dataset is collected with significant camera motions and suitable for representing the scene captured by moving robots' on-board cameras. The dataset consists
of 100 videos obtained from the internet, from which 80 are used for training and 20 for testing. Each video is
captured at 240 frames per second (fps) with a $1280 \times 720$
resolution that we center crop to $256 \times 256$. For each video, we select 80 random $16$-frame-long segments
within the video, therefore $T=16$ in our experiments. The images are turned to gray images and normalized to be within $[0,1]$. Our input and output to the end-to-end model are both the 16-frame video segment. 

\subsubsection{Implementation details}
Our model is implemented in PyTorch. The ratio generation parts are trained with the SGD optimizer (momentum=0.9) at a learning rate of $5\times 10^{-3}$. The video reconstruction network is trained with the Adam optimizer at a learning rate of $5\times 10^{-5}$. The training process is as follows. We first fix the ratio at $8/T$ for all spatial locations and train the video reconstruction network for 100 epochs. After that, we gradually increase $\lambda$ in Eq. (\ref{equ:6}) from $5\times 10^{-3}$ to $0.5$ and jointly train the ratio generation parts and video reconstruction network. For each $\lambda$, we train for about $100$ epochs. For baselines with a fixed compression ratio for all locations, we set the ratio from $1/T$ to $8/T$ and train the video reconstruction network for $100$ epochs in each fixed ratio. Furthermore, we consider both cases where the binary mask, $B$, is used or not used. When $B$ is used, it is randomly initialized and fixed during the experiments.

\subsubsection{Results without binary mask}
\begin{figure}[t]
\centering
\includegraphics[scale=0.65]{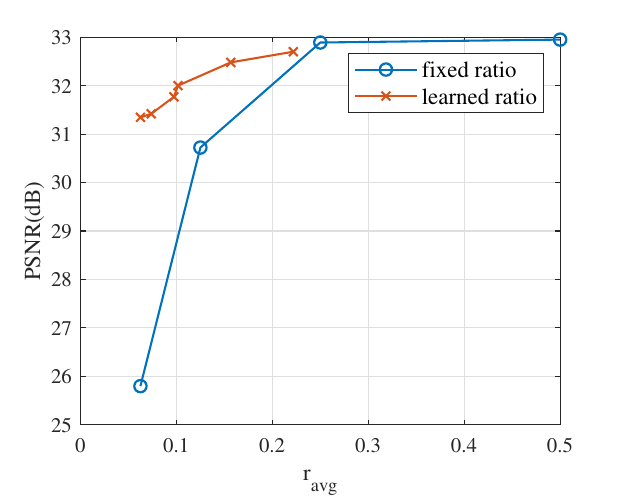}
\caption{The performance comparison between learned spatially-variant ratios methods and fixed ratios methods in video imaging systems without coded aperture.}
\label{fig:sensing}
\end{figure}
We first compare our method with its fixed-ratio version when $B$ is not used. We use the peak signal-to-noise ratio (PSNR), $10\log_{10}(1/\text{MSE}(V,\hat{V})$, as the performance metric. The results are shown in Fig. \ref{fig:sensing}, where $r_{avg}=\frac{1}{HW}\sum_{i=1}^{H}\sum_{j=1}^{W}M_{i,j}$ denotes the average compression ratio for all spatial locations. As shown in Fig. \ref{fig:sensing}, the imaging quality increases along with the growth of $r_{avg}$ for both methods until $r_{avg}$ reaches $0.5\,(8/T)$ where the effects of shot noise and read noise surpass the growth of the number of measurements. From the figure, the proposed method with learned ratios has a significant performance gain over the method with fixed ratio. For example, when $r_{avg}=0.0625 \,(1/T)$, the learned-ratio method has nearly $5$ dB gain over the fixed-ratio method, demonstrating the superiority of learning a compression ratio map in reducing the number of measurement samples. 

\subsubsection{Results with binary mask}
\begin{figure}[t]
\centering
\includegraphics[scale=0.65]{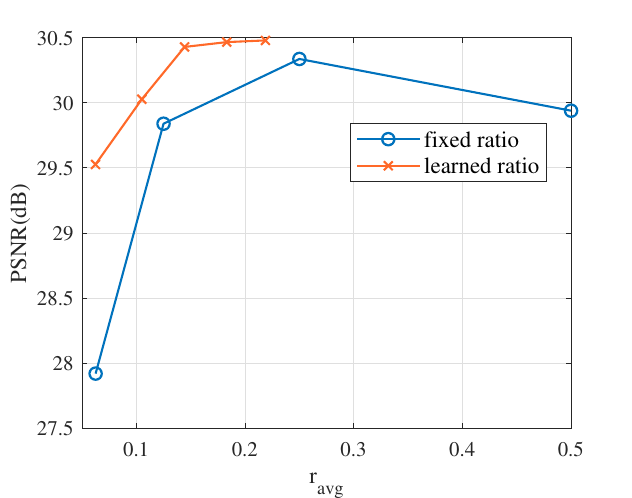}
\caption{The performance comparison between learned spatially-variant ratios methods and fixed ratios methods in video imaging systems with coded aperture.}
\label{fig:sensing2}
\end{figure}
We then consider $B$ and show the performance comparison in Fig. \ref{fig:sensing2}. Due to the usage of binary mask, the system performance of the fixed-ratio method increases when $r_{avg}=0.0625 \,(1/T)$, showing the positive effect of coded apertures. However, as using $B$ will decrease the signal strength, the performance of fixed-ratio method decreases when $r_{avg}$ increases when compared to the same method without $B$ in Fig. \ref{fig:sensing}. From the Figure, the proposed method with learned ratio also has a steady performance gain over the fixed-ratio method, further proving the effectiveness of the developed sensors. 

\subsubsection{Learned compression ratio maps}

\begin{figure}[ht]
  \subfloat[$r_{avg}=0.998$]{
	\begin{minipage}[c][1\width]{
	   0.23\textwidth}
	   \centering
	   \includegraphics[width=1.\textwidth]{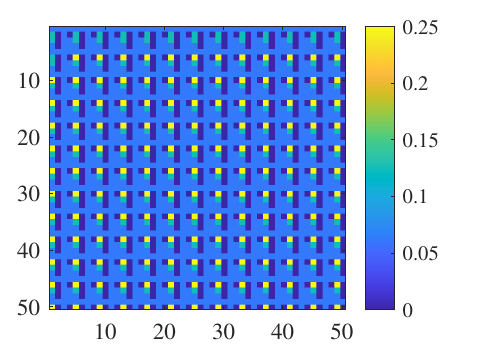}
	\end{minipage}}
 \hfill 	
  \subfloat[$r_{avg}=2.31$]{
	\begin{minipage}[c][1\width]{
	   0.23\textwidth}
	   \centering
	   \includegraphics[width=1.\textwidth]{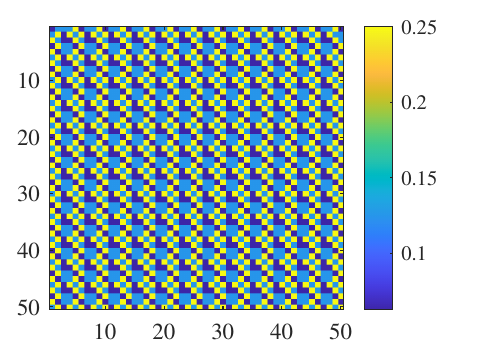}
	\end{minipage}}
 
\caption{The learned ratio maps in a $50\times50$ area of an image when $B$ is used.}
\label{fig:maps}
\end{figure}
We show the learned ratio maps when $B$ is used in Fig. \ref{fig:maps}. As discussed above, there are only five ratio choices from $0$ to $8/T$ and their colors are shown in the color bar. From the figure, the learned ratio maps have fixed patterns for a local area. The patterns are different for different $r_{avg}$. Besides, the learned ratio maps are a mix of low ratios and high ratios so that the benefits from both long exposure and low exposure can be considered. 

\subsubsection{Visual results}
\begin{figure*}[t]
\centering
\includegraphics[scale=0.4]{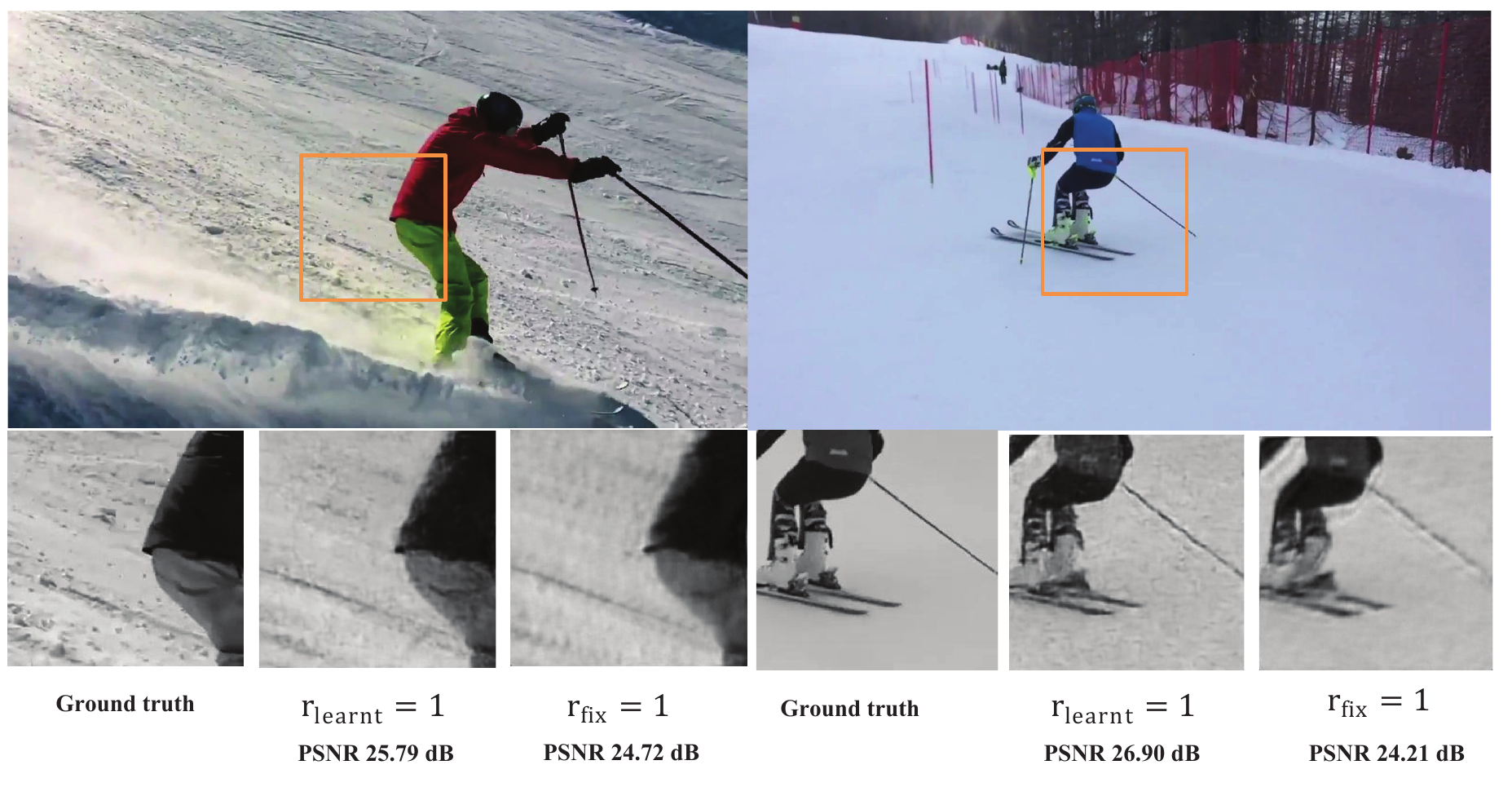}
\caption{Examples of restored video frames from sensing methods with or without learned ratios when compression ratio equals to 1/16. }
\label{fig:vis}
\end{figure*}
Fig. \ref{fig:vis} shows the restored video frames from different methods. We use the fixed-ratio method with $r_{avg}=1$ and the learned-ratio method with $r_{avg}\simeq 1$ in Fig. \ref{fig:sensing2}. From the figure, the frames from learned-ratio method have more texture details and less artifact.
\subsection{Semantic communications for programmable sensors} In this subsection, we will first evaluate semantic communication frameworks based on the task-aware compression and the joint design, respectively. After that, we will compare semantic communications with the existing transmission methods. 

In the following experiments, we first restore the network parameters related to the sensors from pretrained models in the previous subsection and fix them when training the semantic communication frameworks. We use the sensor network with $r_{avg}=0.156$ in Fig. \ref{fig:sensing}.
The experiments are called fixed-sensing experiments. Next, we jointly train the sensing network with the communication parts, which are called joint-sensing experiments.
\subsubsection{Channel condition} 
We assume the sensor data is transmitted through the AWGN channel with $\text{SNR}=10$ dB. 
\subsubsection{Implementation details of different semantic communication frameworks} 
We now describe in more details about the implementations of different semantic communication frameworks.
\begin{itemize}
    \item \textbf{Task-aware compression plus capacity-achieving channel coding (Compr+Cap):} We first convert the sensor data into bitstreams and then assume the transmission of the bitstreams can reach Shannon channel capacity. In the considered channel condition, $\frac{l_{s}}{l_{b}}=0.289$. Note that it is hard to achieve Shannon capacity in real systems, so its performance can only be regarded as an ideal reference for compression methods. To adjust the communication costs, we train the compression network under different $\beta$. During training, we first set $\beta=10^{-7}$ and gradually decrease $\beta$ to increase the bitstream length. The network is first trained with the Adam optimizer at a learning rate of $5\times 10^{-5}$ for about $50$ epochs and then at $5\times 10^{-6}$ for another $50$ epochs under each $\beta$. 
    \item \textbf{Task-aware compression plus LDPC plus QAM (Compr+LDPC):} The bitstreams from task-aware compression methods are transmitted through LDPC channel coding and quadrature amplitude modulation (QAM) modulation. As stated in earlier work \cite{10198383}, when $\text{SNR}=10$ dB for AWGN channels, we should use 16QAM and 2/3 LDPC. In this case, $\frac{l_{s}}{l_{b}}=0.376$.
    \item \textbf{Semantic communications without RAN (SemCom+noRAN):} In earlier task-oriented semantic communications \cite{jankowski2020wireless}, the RAN is not implemented and all the source data is transmitted with the same communication costs. We also delete the RAN in the proposed framework and demonstrate its performance as a reference. To adjust the communication costs, we change the channel dimension of $X$ in Eq. (\ref{equ:9}) from $16$ to $48$. The network is trained for about $50$ epochs at a learning rate of $5\times 10^{-5}$ and then for $50$ epochs at $5\times 10^{-6}$ under each dimension of $X$.
    \item \textbf{Semantic communications (SemCom):} This is the proposed end-to-end semantic communication framework with rate control. To adjust the transmission rate, we train the network under different $\mu$ in Eq. (\ref{equ:12}). The training consists of two steps: we first set $f=4$ for all locations and train the SCE/SCD with all 48 symbols used for about $80$ epochs. After that, we set $\mu=1e-3$ and gradually increase it to decrease communication costs. The RAN is trained with the SGD optimizer and the other parts are trained with the Adam optimizer at a learning rate of $5\times 10^{-5}$ for $50$ epochs and at $5\times 10^{-6}$ for another $50$ epochs under each $\mu$. For low $l_{s}$ situations, we transmit the first $6f$ symbols of $X$ rather than $12f$. Also, exploration strategies are used when sampling $f$. Specifically, the sampling probability is set to $0.6P_{F}+0.4\hat{p}$, where we set $\hat{p}(1)$=$\hat{p}(2)$=$\hat{p}(3)$=$\hat{p}(4)$=$0.25$ to encourage the RAN to explore more on other actions so that the SCE/SCD can perform well on all actions. 
\end{itemize}
\subsubsection{Comparison of semantic communications in fixed-sensing experiments}
\begin{figure}[t]
\centering
\includegraphics[scale=0.65]{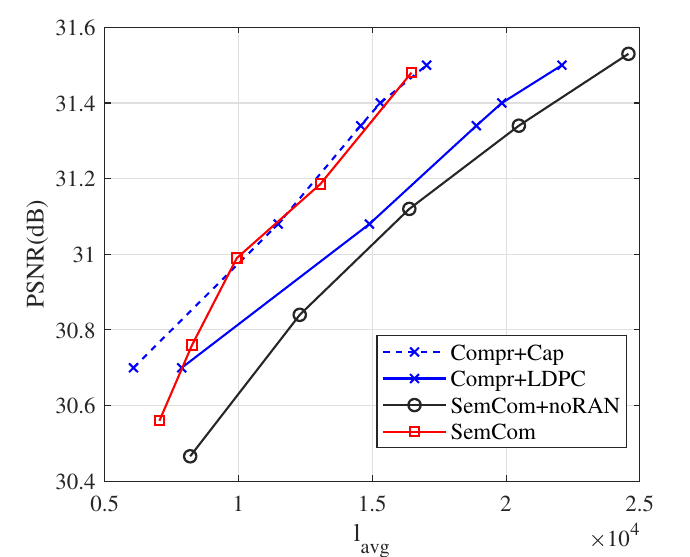}
\caption{The performance comparison among different semantic communication frameworks in fixed-sensing experiments.}
\label{fig:transmission}
\end{figure}
We show the performance comparison of different semantic communication methods in the fixed-sensing experiment in Fig. \ref{fig:transmission}, where $l_{avg}$ denotes the average number of modulated symbols $l_{s}$ used for the video clips in the test dataset. From the figure, the 'Compr+LDPC' performs slightly better than 'SemCom+noRAN'  while the proposed 'SemCom' method outperfroms these methods to a relatively large extend, showing the advantage of jointly designing the channel coding and modulation. It also demonstrates the effectiveness of directly implementing the rate-distortion trade-off on the modulated symbols through the proposed RAN. Furthermore, the proposed 'SemCom' has a similar performance with 'Compr+Cap', showing that semantic communications with joint designs are promising ways to approach Shannon capacity. 

\subsubsection{Comparison of semantic communications in joint-sensing experiments}
\begin{figure}[t]
\centering
\includegraphics[scale=0.65]{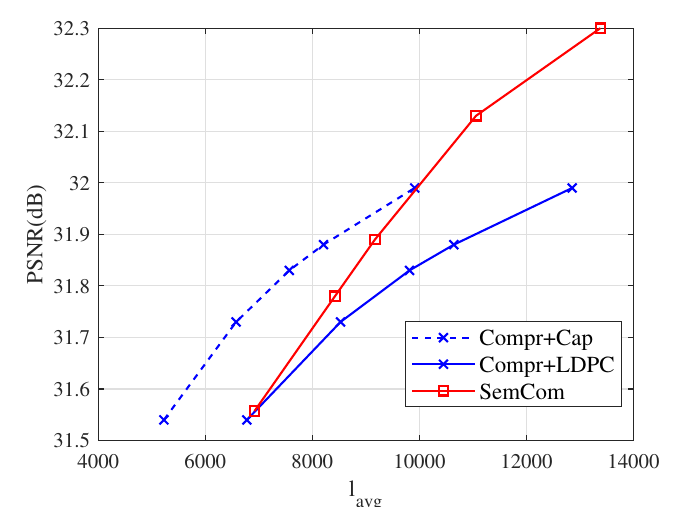}
\caption{The performance comparison among different semantic communication frameworks in joint-sensing experiments.}
\label{fig:transmission2}
\end{figure}
The performance comparison of different semantic communication methods in the joint-sensing experiment is shown in Fig. \ref{fig:transmission2}. From the figure, SemCom performs significantly better than 'Compr+LDPC' and even surpasses the 'Compr+Cap' in large $l_{avg}$ cases, further proving the benefits of joint designs. However, we cannot conclude for sure that semantic communications can surpass Shannon capacity as the performance of 'Compr+Cap' depends on our implementation of task-aware compression methods.

\subsubsection{Implementation of conventional communication methods}
We now explain how the sensor data can be transmitted in conventional communication systems and introduce their implementation details.
\begin{itemize}
\item \textbf{Transmit sensor data by joint source and channel coding (Sensordata+JSCC):} In conventional communication systems, the most straightforward way is to directly transmit the raw sensor data regardless of its usage. To simulate this process, the raw sensor data need to be compressed and channel-coded. Recent works on deep JSCC \cite{bourtsoulatze2019deep} have shown that training a network for joint source and channel coding can perform better than using standardized image compression methods and channel coding methods. Therefore, we follow its design and build a deep network similar to SCE/SCD to transmit the raw sensor. The network is optimized by the mean-squared error (MSE) between original sensor data and reconstructed sensor data.
\item \textbf{Transmit reconstructed video by H.264 video coding plus LDPC plus QAM (Video+H.264+LDPC):} Another choice is to reconstruct the video first via a locally-deployed video reconstruction network and then transmit the reconstructed video. In this way, the computational-costive reconstruction network need to be run at the transmitter. To transmit the video, we use H.264 \cite{wiegand2003overview} for video source coding, LDPC for channel coding, and QAM for modulation. 
\end{itemize}
\subsubsection{Comparison of semantic communications with conventional communication systems}
\begin{figure}[t]
\centering
\includegraphics[scale=0.65]{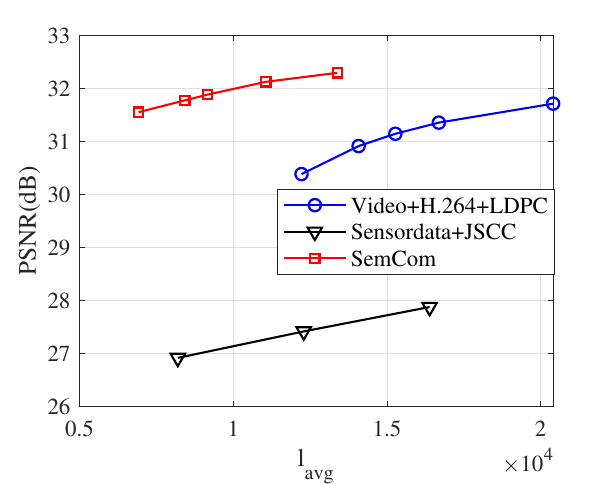}
\caption{The performance comparison among different transmission methods.}
\label{fig:transmission3}
\end{figure}
The performance comparison between semantic communications and conventional communication systems is shown in Fig. \ref{fig:transmission3}. From the figure, Sensordata+JSCC performs the worst. This is because the important data for video reconstruction networks does not get targeted protection during transmission. Video+H.264+LDPC performs better than Sensordata+JSCC as the videos have been reconstructed at the transmitter side. However, this method is still not as effective as SemCom, which shows we can achieve the efficient transmission of sensor data without running time-consuming and resource-costive reconstruction algorithm at the transmitter side. 

\section{Conclusion}
\label{sec:conclusion}
In this work, we propose a novel video imaging system with learned compression ratios. We also propose a semantic communication framework for programmable sensors with task-oriented reconstruction algorithms. From the perspective of algorithm development, we show that by combining the policy-gradient reinforcement learning and supervised learning, we can achieve the explicit (compression or transmission) rate-distortion trade-off in different cases. The proposed training pipelines can be extended to many other applications.  
\bibliographystyle{IEEEtran}
\bibliography{ref}
\small
\end{document}